\documentclass[12pt]{iopart}
\usepackage[]{indentfirst}
\usepackage{amssymb}

\begin{document}

\begin{titlepage}

\title{Quantum Singularity of Levi-Civita Spacetimes}

\author{D A Konkowski\dag, T M  Helliwell\ddag,   and C Wieland\ddag}

\address{\dag Department of Mathematics, U.S. Naval Academy, Annapolis,
Maryland, 21402 U.S.A.}

\address{\ddag Department of Physics, Harvey Mudd College,
Claremont, California 91711 USA}

\eads{\mailto{dak@usna.edu}, \mailto{T\_Helliwell@HMC.edu}}

\begin{abstract}
        Quantum singularities in general relativistic spacetimes are
        determined by the behavior of quantum test particles. A static
        spacetime is quantum mechanically singular if the spatial
        portion of the wave operator is not essentially self-adjoint.
        Here Weyl's limit point-limit circle criterion is used to
        determine whether a wave operator is essentially self-adjoint.
        This test is then applied to scalar wave packets in
        Levi-Civita spacetimes to help elucidate the physical
        properties of the spacetimes in terms of their metric parameters.
\end{abstract}


\maketitle

\end{titlepage}

 \section{Introduction}
 In classical general relativity, singularities are not part of the
spacetime;
 they are boundary points indicated by incomplete geodesics or
 incomplete curves of bounded acceleration in a maximal
 spacetime (see, e.g. \cite{HE, ES}). So at least for timelike
 and null geodesics, this incompleteness can be considered as the
 abrupt ending of classical particle paths.

 \par What happens if, instead of classical particles, one uses quantum
 mechanical particles to identify singularities? This is the question G.
 Horowitz and D. Marolf \cite{HM} set out to answer following earlier
 work by R. Wald \cite{Wald}. They call a spacetime quantum
 mechanically nonsingular if the evolution of a test scalar wave
 packet in the spacetime, representing a quantum mechanical particle,
 is uniquely determined by the initial wave
 packet, without having to place arbitrary boundary conditions at the
 classical singularity. Their construction is restricted to
 static spacetimes.

 \par Mathematically, this evolution is related to properties of a
 quantum mechanical operator. A static spacetime is quantum mechanically
 singular \cite{HM} if the spatial portion of the Klein-Gordon wave
operator is not
 essentially self-adjoint \cite{RS} \cite{Rich}. If this operator is not
 essentially self-adjoint then the evolution of a test scalar wave
 packet is not determined uniquely by the initial wave packet;
 boundary conditions at the classical singularity are needed to ``pick
 out'' the correct wave function and thus one needs to add information
 that is not already present in the wave operator, spacetime metric,
 and manifold. This will be described further in the paragraphs below.

 \par Horowitz and Marolf restrict the wave operator to the Klein-Gordon
 operator, but it has been shown \cite{HKA} that Maxwell and Dirac operators
 may also be used with equivalent generic results. Throughout this paper
 we will restrict the analysis to Klein-Gordon operators.

 \par A relativistic quantum particle with mass $M$ can be described
 by a positive frequency solution to the Klein-Gordon equation

 \begin{equation}
     \frac{\partial ^{2} \Psi}{\partial t^{2}} = -A \Psi
     \label{eq:1}
 \end{equation}

\noindent in a static spacetime \cite{HM}. The spatial Klein-Gordon
operator $A$
is

\begin{equation}
    A \equiv -V D^{i}(V D_{i}) + V^{2} M^{2}
    \label{eq:2}
\end{equation}

\noindent where $V^{2} = - \xi_{\nu}\xi^{\nu}$ (here $\xi^{0}$ is the
timelike
Killing field) and $D_{i}$ is the spatial covariant derivative on the
static slice $\Sigma$. The appropriate Hilbert space $H$ is
$\mathcal{L}^{2}(\Sigma)$, the space of square integrable functions on
$\Sigma$.
This choice agrees with that of Horowitz and Marolf \cite{HM}.  The
volume element used to define $H$ is $V^{-1}$ times the natural volume
element on $\Sigma$. We do
not choose the first Sobolev norm $H^{1}$ proposed by Ishibashi and
Hosoya \cite{IH}; that choice is related to Dirichlet boundary
conditions at the singularity \cite{Wald}. If we initially define
the domain of $A$ to be $C_{0}^{\infty}$, $A$ is a real positive
symmetric operator and self-adjoint extensions always exist \cite{RS}.
If there is one unique self-adjoint extension $A_{E}$, then $A$ is
essentially self-adjoint \cite{RS}. In this case the Klein-Gordon
equation for a free relativistic particle takes the form \cite{HM}

\begin{equation}
    i \frac{\partial \Psi}{\partial t} = (A_{E})^{1/2} \Psi
    \label{eq:3}
\end{equation}

\noindent with

\begin{equation}
   \Psi (t) = \exp ( -i t (A_{E}^{1/2}) \Psi (0).
    \label{eq:4}
\end{equation}

\noindent Equations (\ref{eq:3}) and (\ref{eq:4}) are ambiguous if $A$ is
not essentially
self-adjoint. This fact led Horowitz and Marolf \cite{HM} to define
quantum-mechanically singular spacetimes as those in which $A$ is not
essentially self-adjoint.

\par In their paper Horowitz and Marolf test several classically
singular spacetimes to determine whether they are quantum mechanically
singular as well. They find that Reissner-Nordstr\"{o}m, negative mass
Schwarzschild, and the $2D$ cone remain singular when probed by
quantum scalar test particles; however, certain orbifolds, extreme
Kaluza-Klein black holes, the $D=5$ fundamental string, and a few other
examples are nonsingular. Helliwell and Konkowski \cite{HK} and
Helliwell, Konkowski and Arndt \cite{HKA} show that a broad class of
quasiregular spacetimes (spacetimes with quasiregular singularities)
are quantum mechanically singular. This work was an extension of
earlier work by Kay and Studer \cite{KS} which showed that the $2D$ cone and
a $4D$ idealized cosmic string are not essentially self-adjoint.

\par The work of Horowitz and Marolf \cite{HM} and subsequent work by
Helliwell, Konkowski and Arndt \cite{HKA} test for essential
self-adjointness of the spatial
operator $A$ by using a von Neumann criterion \cite{VN, weyl} . This method
involves studying solutions to

\begin{equation}
    A \Psi = \pm i \Psi
    \label{eq:5}
\end{equation}

\noindent and finding the number of solutions that belong to
$\mathcal{L}^{2}(\Sigma)$
for each sign of $i$. This determines the von Neumann deficiency indices
which indicate whether the operator is essentially self-adjoint or
whether it has self-adjoint extensions, and how many self-adjoint
extensions it has \footnote{For an
introduction to the mathematics of self-adjoint operators and von
Neumann deficiency indices, see \cite{HKA} which is based on \cite{RS, IH,
BFV}} . If
the deficiency indices are $(0,0)$, so that no solutions are square
integrable, then the operator is essentially self-adjoint and so has
a unique self-adjoint extension. The wave behavior described by the
operator is uniquely determined for all time by the operator, the
spacetime metric, and the manifold. No additional information in the
form of boundary conditions need be added.

\par In this paper we use an alternative test for quantum
singularity based upon the limit point-limit circle criterion for
essential self-adjointness of operators. That is, we use a theorem of Weyl's
\cite{RS} to relate the essential self-adjointness of the operator via
the von Neumann deficiency indices to the ``potential'' which
determines the behavior of the scalar wave packet. The effect is
determined by the limit point-limit circle criterion
which we will discuss in Section 3. After developing this technique in
the context of determining the quantum singularity nature of a general
spacetime, we
study the so-called Levi-Civita spacetimes in particular.

\par Levi-Civita spacetimes are static, primarily cylindrically symmetric,
spacetimes that are classically singular at ``$r=0$'' unless the
metric is Minkowski or Minkowski in accelerated coordinates. They are
used to model infinite line masses and idealized cosmic strings.
There has been some controversy in the literature about the physical
relevance of certain parameters; this will be reviewed in Section 2.
The quantum singularity or nonsingularity of $r=0$ will be studied in
Section 4 (following the mathematical background in Section 3).  Finally
in Section 5
we use the quantum singularity results
for certain parameter values to glean what information we can about the
physical
interpretation of the corresponding spacetimes.

\section{Levi-Civita Spacetimes}

The metric in Levi-Civita spacetimes \cite{LC} has the form

\begin{equation}
    ds^{2} = r^{4 \sigma}dt^{2} - r^{8 \sigma^{2} - 4\sigma}(dr^{2}+
    dz^{2}) - \frac{r^{2 - 4 \sigma}}{C^{2}} d\theta^{2}
    \label{eq:6}
\end{equation}

\noindent where $\sigma$ and $C$ are real numbers ($C>0$).  It is
straightforward to show that if $\sigma$ is replaced by $1/4\sigma$
in the metric, the spacetime is unchanged, although a coordinate
transformation is required to make this obvious.  Therefore we
need only study the range $-\frac{1}{2}\le \sigma\le\frac{1}{2}$.
Neither of the parameters $\sigma$ and $C$ can
be removed by a coordinate transformation \cite{bonnor,
HSTW, HRS}.  Computation of the Kretschmann scalar

\begin{equation}
    R_{\mu \nu \sigma \tau}R^{\mu \nu \sigma \tau} = \frac{64
    \sigma^{2} (2\sigma -1)^{2}}{(4 \sigma^{2} -2\sigma +1)^{3} r^{4}}
    \label{eq:7}
\end{equation}

\noindent shows that $r=0$ is a scalar curvature singularity for all
$\sigma $
except $\sigma = 0, 1/2$.

\par If $\sigma = 0$, the metric is

\begin{equation}
    ds^{2} = -dt^{2} + dr^{2} +dz^{2} + \frac{r^{2}}{C^{2}} d \theta^{2}.
    \label{eq:8}
\end{equation}

\noindent If $C=1$, this is simply Minkowski spacetime in
cylindrical coordinates. It thus seems reasonable to choose the
coordinate ranges

\begin{equation}
    -\infty < t < \infty,\,  -\infty < z < \infty,\,
    0 < r < \infty, \,  0  \le \theta \le 2 \pi,
    \label{eq:9}
\end{equation}

\noindent with $0$ and $2\pi$ identified. If $C \not= 1$, equation
\ref{eq:8} is the
metric for an idealized cosmic string. There is a quasiregular
(``disclination'') singularity at $r=0$ (see, e.g., \cite{HKA} for a
discussion). This is a
topological singularity, not a curvature singularity. It is thus
reasonable to identify the parameter $C$ in the general metric,
equation (6), with a topological property of the spacetime, that is,
the deficit angle.

\par The sigma parameter is less easily interpreted \cite{bonnor,
HSTW, HRS}. For certain values one can interpret these
spacetimes as those of infinite line masses. This
interpretation is clearest for $|\sigma|\ll 1$. The most physically
realistic range is $0<\sigma<1/4$,  where $\sigma$
represents the mass per unit length and timelike circular orbits exist.
As $\sigma$ increases from 1/4 to 1/2 the interpretation is more
difficult \cite{bonnor, HSTW, HRS}. In this range, the Kretschmann
scalar decreases with increasing $\sigma$, which seems to suggest that the
gravitational field is becoming weaker. But as several authors comment
\cite{bonnor, HSTW, HRS}, the Kretschmann scalar may not be a good
indicator of gravitational
strength; the acceleration of test particles, increasing with
increasing $\sigma$ in the interval $(1/4, 1/2)$, may be a better
indicator.  Interior solutions matching to exterior Levi-Civita spacetimes
exist for the entire range $0<\sigma<1/2$. Bonnor \cite{bonnor}
proposes to interpret Levi-Civita as the spacetime generated by a
cylinder whose radius increases with increasing $\sigma$ and tends to
infinity as $\sigma$ tends to $1/2$. Thus the entire range
$0<\sigma<1/2$ can be taken to represent an ``infinite line mass''.

\par The $\sigma = 1/2$ case is more problematic. Bonnor's analysis
suggests (but does not prove) that when $\sigma = 1/2$ the cylinder
becomes a plane, an interpretation first put forward by Gautreau and
Hoffmann \cite{GH} based on different considerations. In fact, the $\sigma
=1/2$ metric

\begin{equation}
    ds^{2} = -r^{2} dt^{2} + dr^{2} + dz^{2} + \frac{1}{C^{2}}
    d\theta^{2}
    \label{eq:10}
\end{equation}

\noindent is flat. We can transform to Minkowski coordinates

\begin{equation}
    ds^{2} = -d{\bar t}^{2} + d{\bar x}^{2} + d{\bar y}^{2} + d{\bar
    z}^{2}
    \label{eq:11}
\end{equation}

\noindent if we let
${\bar t} = r \sinh t$, ${\bar x} = r \cosh t$, ${\bar y} = \theta/C$,
and ${\bar z} = z$. Here we can obviously let the ${\bar y}$ coordinate range
from $-\infty$ to $\infty$. This is flat spacetime described from the
view of an accelerating frame of reference. This seems to support the
interpretation of the $\sigma = 1/2$ case as a planar source producing
flat spacetime described by a uniformly accelerating observer
\cite{bonnor}. In other words, one can interpret it as the spacetime of a
gravitational field produced by an
infinite planar sheet of positive mass density.

\par On the other hand, one can interpret the metric with $\sigma = -1/2$
as the
gravitational field produced by an infinite sheet of negative mass
density (an idea also first proposed by Gautreau and Hoffmann
\cite{GH}). However, this case is also problematic \cite{bonnor, HSTW,
HRS}. The metric

\begin{equation}
    ds^{2} = - \frac{1}{r^{2}} dt^{2} + r^{2} (dr^{2} + dz^{2} +
    \frac{1}{C^{2}} d \theta^{2})
       \label{eq:12}
\end{equation}

\noindent is not flat and it has a curvature singularity at $r=0$. It
has planar symmetry with one additional extra Killing vector beyond
those in the general
Levi-Civita spacetimes. The interpretation of this
spacetime as one caused by an infinite plane of negative mass density
comes from the fact that test particles are repelled from the $r=0$
plane and the fact that ``the Gaussian curvature of spacelike
`eigensurfaces' is
zero'' \cite{GH}. This is a reasonable interpretation; the
``problem'' comes from interpreting both $\sigma = 1/2$ and $\sigma
=-1/2$ as planes of infinite (positive/negative) mass density as one
is singularity-free and the other has a curvature singularity
\cite{HSTW, HRS}.

\par We will now study the quantum singularity properties of these
spacetimes for various parameter values in the following sections. We
begin by introducing the mathematics necessary to discuss essentially
self-adjoint operators and quantum singularities.

\section{Weyl's Limit Point - Limit Circle Criterion}

\par  A particularly convenient way to establish essential
self-adjointness in the spatial operator of the Klein-Gordon equation
is to use the concepts of limit circle and limit point behavior.\footnote
{This section is based on {\bf Appendix to X.1} in Reed and
Simon \cite{RS}}  The
approach is as follows.  The Klein-Gordon equation for Levi-Civita
spacetimes can be separated in the coordinates $t, r, \theta, z$.
Only the radial equation is non-trivial.  With changes in both
dependent and independent variables, the radial equation can be
written as a one-dimensional Schr\"{o}dinger equation

\begin{equation}
    H\Psi(x) = E\Psi(x)
    \label{eq:13}
\end{equation}

\noindent where $x \in (0,\infty )$ and the operator  $H = - d^{2}/dx^{2}
+ V(x)$.

\newtheorem{Definition}{Definition}
\begin{Definition} The potential $V(x)$ is in the limit circle case
at $x = 0$  if for some, and therefore for all $E$, {\it all}
solutions of equation (\ref{eq:13}) are square integrable at zero.  If
$V(x)$  is not
in the limit circle case, it is in the limit point case.
\end{Definition}

\par  A similar definition pertains to $x=\infty$.  The potential
$V(x)$ is in the limit circle case at $x=\infty$ if all solutions of
equation (\ref{eq:13}) are square integrable at infinity; otherwise,
$V(x)$ is in
the limit point case at infinity.

\par There are of course two linearly independent solutions of the
Schr\"{o}dinger equation for given $E$.  If $V(x)$ is in the limit circle
case at zero, both solutions are $\mathcal{L}^{2}$ at zero, so all linear
combinations are $\mathcal{L}^{2}$ as well.  We would therefore need a
boundary
condition at $x=0$ to establish a unique solution.  If $V(x)$ is in
the limit {\it point} case, the $\mathcal{L}^{2}$ requirement eliminates
one of
the solutions, leaving a unique solution without the need of
establishing a boundary condition at $x=0$. This is the whole idea of
testing for quantum singularities; there is no singularity if the
solution is unique, as it is in the limit point case.  The critical
theorem is due to Weyl \cite{RS}.

\newtheorem{theorem}{Theorem}
\begin{theorem}[The Weyl limit point-limit circle criterion.] If
$V(x)$ is a continuous real-valued function on $(0, \infty)$, then
$H = - d^{2}/dx^{2} + V(x)$ is essentially self-adjoint on
$C_{0}^{\infty}(0, \infty)$ if and only if $V(x)$ is in the limit
point case at both zero and infinity.
\end{theorem}

\par The following theorem can be used to establish the limit circle-limit
point behavior at infinity \cite{RS}.

\newtheorem{theorem1}[theorem]{Theorem}
\begin{theorem1}[Theorem X.8 of Reed and Simon \cite{RS}.] If $V(x)$ is
continuous and real-valued on $(0, \infty)$, then $V(x)$ is in the
limit point case at infinity if there exists a {\em positive}
differentiable function $M(x)$ so that
\begin{enumerate}
    \item[(i)] $V(x) \ge - M(x)$
    \item[(ii)] $\int_{1}^{\infty} [M(x)]^{-1/2} dx = \infty$
    \item[(iii)] $M'(x)/M^{3/2}(x)$ is bounded near $\infty$.
\end{enumerate}
Then $V(x)$ is in the limit point case (complete) at $\infty$.
\end{theorem1}

A sufficient choice of the $M(x)$ function for our purposes is the
power law function $M(x) = c x^{2}$ where $c > 0$. Then {\it (ii)} and
{\it (iii)} are satisfied, so if $V(x) \ge -c x^{2}$, $V(x)$ is in the
limit point case at infinity.

\par A theorem useful near zero is the following.

\newtheorem{theorem2}[theorem]{Theorem}
\begin{theorem2} [Theorem X.10 of Reed and Simon \cite{RS}.]
Let $V(x)$ be continuous and {\it positive} near zero.
If $V(x)\ge\frac{3}{4} x^{-2}$ near zero then $V(x)$ is in the limit
point case.  If for some $\epsilon > 0$,
$V(x)\le(\frac{3}{4}-\epsilon)x^{-2}$ near zero, then $V(x)$ is in the
limit circle case.
\end{theorem2}

\noindent These results can now be used to help test for quantum
singularities in the Levi-Civita spacetimes.

\section{Limit Point - Limit Circle properties of Levi-Civita Spacetimes}

\par The Klein-Gordon equation for a scalar particle of mass $M$ is

\begin{equation}
     \Box \Phi \equiv g^{\mu \nu} \Phi_{\,\mu \nu} +
     \frac{1}{\sqrt{g}}(\sqrt{g}g^{\mu \nu}),_{\nu} \Phi_{\,\mu} = M^{2}
     \Phi ;
     \label{eq:14}
\end{equation}

\noindent for the Levi-Civita metrics it is

\begin{equation}
    \fl -r^{-4 \sigma} \Phi,_{tt} + r^{8 \sigma^{2} + 4 \sigma}
    (\Phi,_{rr} + \Phi,_{zz}) +C^{2} r^{4 \sigma - 2}
    \Phi,_{\theta\theta} +
    r^{-8 \sigma^{2} + 4\sigma - 1} \Phi,_{r} = M^{2} \Phi.
    \label{eq:15}
\end{equation}

\noindent The equation separates using $\Phi = e^{-i \omega
t}e^{ikz}e^{im \theta} R(r)$, where $\omega, k$ are continuous and $m=0,
\pm 1, \pm 2, \cdots$. The resulting radial equation is

\begin{equation}
    \frac {d^{2} R}{d r^{2}} + \frac{1}{r} \frac{d R}{d r} + \left[
    \omega^{2} r^{8 \sigma^{2}- 8 \sigma} - k^{2} - M^{2} r^{8
    \sigma^{2} - 4 \sigma}-m^{2} C^{2} r^{8 \sigma^{2} - 2} \right]
    R = 0.
    \label{eq:16}
\end{equation}

\noindent As described in Section 1, square integrability is judged by
evaluating
the integral

\begin{equation}
    I = \int dr \sqrt{g_{3}/g_{00}} R^{*} R = (1/C) \int dr  r^{8
    \sigma^{2} - 8 \sigma +1} R^{*} R
    \label{eq:17}
\end{equation}

\noindent where $g_{3}$ is the negative determinant of the 3-space portion of
the metric tensor \cite{Wald, HM}.  The resulting integral is
invariant under coordinate transformations.

\par To test for limit point-limit circle behavior one must
rewrite the radial equation in the one-dimensional
Schr\"{o}dinger-equation form

\begin{equation}
    \frac{d^{2} \Psi (x)}{dx^{2}} + \left[E - V(x) \right] \Psi = 0,
    \label{eq:18}
\end{equation}

\noindent for which square integrability is judged by the integral $I =
\int dx
\Psi^{*} \Psi$.  Except for the special case $\sigma = 1/2$, which we
return to later, the
appropriate transformations of dependent and independent variables are
$R = (1/\sqrt{x}) \Psi$ and $r = ( \alpha C x^{2})^{1/(2 \alpha)}$
where $\alpha \equiv (2 \sigma - 1)^{2}$.

\par Equation (16) transforms to equation (18) if we let $E = C
\omega^{2}/ \alpha$ and
\begin{eqnarray}
    V(x) & = & (C k^{2}/ \alpha) ( \alpha C x^{2})^{(- \alpha + 1)/
    \alpha} + (C M^{2}/ \alpha) ( \alpha C x^{2})^{2 \sigma/
    \alpha} \nonumber\\
      & + & (m^{2} C^{3}/ \alpha) (\alpha C x^{2})^{(4 \sigma - 1)/
    \alpha} - 1/(4 x^{2}).
    \label{eq:19}
\end{eqnarray}

\noindent In the limit $x \rightarrow \infty$ the final term vanishes,
leaving
a non-negative $V(x)$. Therefore the solutions are limit point at
infinity for all $\sigma$ ($\sigma \not= 1/2$), since as $x \rightarrow
\infty$, $V(x) > - c x^{2}$ for any positive constant $c$ \cite{RS,
Rich}.

\par It remains to test the solutions in the limit $x \rightarrow 0$.
This we do by identifying the important terms in $V(x)$. The final term
diverges as $x^{-2}$. The first term in $V(x)$ diverges less quickly for
any finite  $\sigma$. The second term $\sim x^{4 \sigma/ \alpha}$,
whose exponent has the minimum value $-1/2$ at $\sigma = -1/2$. The
third term $\sim x^{2( 4 \sigma - 1)/ \alpha}$, whose exponent has the
minimum value -2 at $\sigma = 0$.

\par Therefore unless $\sigma = 0$ (or $\sigma = 1/2$, which we
treat later), the equation has the form

\begin{equation}
    \frac{d^{2} \Psi}{dx^{2}} + \frac{1}{4 x^{2}} \Psi = 0
    \label{eq:20}
\end{equation}

\noindent as $x \rightarrow 0$.  By Theorem 3 the solutions
should exhibit limit circle behavior; in fact, one solution is $\Psi_{1} =
x^{1/2}$
and the other is $\Psi_{2} = \Psi_{1} \int^{x} dx/
\Psi_{1}^{2} = x^{1/2} \ln (x)$. Both are $\mathcal{L}^{2}$
near $x = 0$, so are indeed limit circle.

\par If $\sigma =0$, the equation has the form
\begin{equation}
    \frac{d^{2} \Psi}{d x^{2}} - \frac{m^{2} C^{2} - 1/4}{x^{2}}
    \Psi = 0
    \label{eq:21}
\end{equation}

\noindent as $x \rightarrow 0$. One solution is $\Psi_{1} = x^{1/2 + |m| C}$;
the other is $\Psi_{2} = x^{1/2} \ln (x)$ (if $m = 0$), or $
\Psi_{2} = x^{1/2 - |m| C}$ (if $m \not= 0$). Both solutions are
limit circle if $|m| C < 1$, and limit point if $|m| C \ge 1$.

\par In the special case $\sigma = 0,\, C = 1$, the spacetime is
Minkowski with no classical singularity at $r =0$. In that case $x =
r$ and $R_{1} = 1, R_{2} = \ln (r)$. The solution $R_{2} = \ln (r)$ is
unacceptable in this case because it diverges at $r =0$ even though $r
=0$ is a
regular hypersurface in the spacetime.  \footnote {We assumed at the
outset that
the domain of the spatial operator consists
of smooth functions with compact support away from the origin.
For Minkowski space the points $r = 0$ are arbitrary, so it is
unphysical to permit a function that diverges at  $r = 0$  and nowhere else.}
If $\sigma = 0,\, C \neq 1$ the spacetime
corresponds to a linear idealized cosmic string with a quasiregular
singularity at $r = 0$.

\par For the special case $\sigma = 1/2$ the radial Klein-Gordon
equation is

\begin{equation}
    \frac{d^{2} R}{dr^{2}} + \frac{1}{r} \frac{d R}{dr} +\left[ -(
    k^{2} + M^{2} + m^{2} C^{2}) + \omega^{2}/ r^{2} \right] R = 0.
    \label{eq:22}
\end{equation}

\noindent The transformation $R(r) = \Psi (x)$ and $r = \exp (C x)$ puts the
equation in the form

\begin{equation}
    \frac{d^{2} \Psi}{dx^{2}} + \left[ E - V(x) \right] \Psi = 0
    \label{eq:23}
\end{equation}

\noindent where $E = C^{2} \omega^{2}$ and

\begin{equation}
    V(x) = C^{2} ( k^{2} + M^{2} + m^{2} C^{2}) \exp (2Cx)
    \label{eq:24}
\end{equation}

\noindent with  $\int dr \sqrt{g_{3}/g_{00}} R^{*} R = \int dx \Psi^{*} \Psi$.

\par In the transformation $r = 0$ corresponds to $x = - \infty$, while
$r = \infty $ corresponds to $x = \infty$. So in this case we must
determine the limit point-limit circle behavior at $x = \pm \infty $.
In this case $V(x) > -c x^{2}$ at both $\pm \infty $, for any
positive $c$. Therefore the solutions are limit point at both ends.
It is interesting that the $\sigma = 1/2$ case is a special case
which must be studied on the interval $(-\infty, \infty)$.  This result seems
to support Bonnor's description of the $\sigma = 1/2$ case as a
cylinder whose radius tends to $\infty$ as $\sigma$ tends to $1/2$.

\section{Conclusions}

\par In the preceding section we showed that the Klein-Gordon operator
is limit point at infinity for the whole class of Levi-Civita
spacetimes. If $\sigma$ is neither zero nor one-half, the
Klein-Gordon operator is limit circle at $r = 0$ and thus not
essentially self-adjoint,  so all $\sigma \neq
0, \sigma \neq 1/2$ Levi-Civita spacetimes are quantum mechanically
singular as well as being classically singular.

\par If $\sigma = 0$ and $C = 1$, the spacetime is simply Minkowski
space. One of the two solutions of the radial Klein-Gordon equation
can be rejected because it diverges at a regular point ($r=0$) of the
spacetime. The operator is therefore limit point at $r=0$, and
so Minkowski spacetime is quantum
mechanically nonsingular (a well known fact, repeated here for
completeness).

par If $\sigma = 0$ and $C \neq 1$, the spacetime is the conical
spacetime corresponding to an idealized cosmic string. The operator
is limit circle if $|m| C < 1$ and limit point if $|m| C \ge 1$.
Therefore the cosmic string spacetimes are quantum mechanically
singular for azimuthal quantum number $m$ such that $|m| C < 1$ and
nonsingular if  $|m| C \ge 1$. If arbitrary values of $m$ are allowed,
these
spacetimes are quantum mechanically singular in agreement with earlier
results \cite {HK, HKA, KS}. Recall that these spacetimes are also
classically singular with a quasiregular (``disclination'')
singularity at $r=0$.

\par If $\sigma = 1/2$ the classical spacetime is flat and without a
classical singularity. We have shown that this spacetime is quantum
mechanically nonsingular as well for massive scalar particles obeying the
Klein-Gordon equation, since the operator is limit point at both ends
and thus essentially self-adjoint.

\par Thus for the Levi-Civita spacetimes, all that are classically
singular are also quantum mechanically singular, and all that are
classically nonsingular ($\sigma = 0,\, C = 1$, and $\sigma = 1/2$) are
also quantum mechanically nonsingular.  The classically and
quantum-mechanically nonsingular spacetimes correspond to isolated
values of $\sigma$, so that (for example) even though the spacetime
$\sigma = 0,\, C = 1$ is nonsingular, the spacetimes with  $\sigma
\to 0,\, C = 1$ are singular.  The only discrepency between
classical and quantum singularities are for the $\sigma =0,\, C \neq 1$
modes with $|m| C \ge 1$, which are quantum mechanically nonsingular
in a classically singular spacetime. The physical reason is that the
wavefunction for large values of $m$ in a flat space with a quasiregular
singularity at $r=0$ are unable to detect the presence of the
singularity because of the repulsive centrifugal potential
$m^{2}C^{2}/x^{2}$ in equation (21).  The limit point/limit circle
criterion, together with the theorems that connect it to the effective
potential in a Schrodinger equation, provide physical insight into
when quantum singularities are prevented from occurring by potential
barriers.

\ack We thank David Clarke, Jack Harrison and Cassidi Reese for useful
conversations.
One of us (DAK) was partially funded by NSF grant PHY99-88607 to
the U.S. Naval Academy. She also thanks Queen Mary, University of London,
where some of this work was carried out.

\end{document}